\documentclass[12pt]{article}
\usepackage{epsf}

\def\lrel{{\cal L}_{\rm relax}}
\def\la{\langle}
\def\ra{\rangle}

\def\dt{\delta t}
\def\phia{\phi^{(1)}}
\title{ A Wave Function approach \\ to dissipative processes}
\author{ Yvan Castin and Jean Dalibard
         \\Laboratoire de Spectroscopie Hertzienne de l'ENS
         \thanks{Unit\'e de recherche de l'Ecole Normale
             Sup\'erieure et de l'Universit\'e Paris 6,
             associ\'ee au CNRS.}
         \\24 rue Lhomond, F-75231 Paris Cedex 05, France
      \\ and Klaus M\o lmer
        \\ Max-Planck-Institut f\"ur Quantenoptik,
        \\ W-8046 Garching, Germany
        \thanks{Permanent address: Institute of Physics and Astronomy,
          Aarhus University, DK-8000 Aarhus C, Denmark}
}
\date{}

\begin{document}

\maketitle

\setlength{\baselineskip}{16pt}
\vskip 15mm

{\bf Abstract:} {{\sl We present a wave function approach to study the evolution of a small system
when it is coupled to a large reservoir. Fluctuations and dissipation originate in this approach from
quantum jumps occurring randomly during the time evolution of the system. This approach can be applied
to a wide class of relaxation operators in the Markovian regime, and it is equivalent to the standard
Master Equation approach.}\\ \\\textrm{Published in AIP Conference Proceedings 275, Thirteenth
International Conference on Atomic Physics, Munich, Germany 1992; Editors: H. Walther, T. W. H\"ansch,
and B. Neizert.}}

\vskip 15mm

The problem of dissipation plays a central role in Atomic Physics
and Quantum Optics. The simplest example is the phenomenon of
spontaneous emission, where the coupling between an atom and the
ensemble of modes of the quantized electromagnetic field gives a
finite lifetime to all excited atomic levels. Usually the
dissipative coupling between a small system and a large reservoir
can be treated by a master equation approach
\cite{Louisell,Haake,Cohen,Gardiner}; one writes a linear equation
for the time evolution of the reduced system density matrix
$\rho_S={\rm Tr_{res}} (\rho)$, trace over the reservoir variables
of the total density matrix. If we denote  the hamiltonian for the
system $H_S$, this equation can be written:
\begin{equation}
\dot \rho_S={i\over\hbar}[\rho_S,H_S] + \lrel(\rho_S)  \ .
\label{relax}
\end{equation}
In (\ref{relax}), $\lrel$ is the relaxation superoperator, acting on
the density operator $\rho_S$. It is assumed here to be local in
time, which means that $\dot \rho_S(t)$ depends only on $\rho_S$ at
the same time (Markov approximation).  All the system dynamics can
be deduced from (\ref{relax}). One can calculate one time average
values of a system operator $A$: $a(t)=\la A\ra (t)={\rm
Tr}(\rho_S(t)A)$, and also, using the quantum regression theorem
\cite{Lax}, multi-time correlation functions such as $\la
A(t+\tau)B(t)\ra$.

We present here an alternative treatment  based on a Monte-Carlo
evolution of wave functions of the small system (MCWF)
\cite{Dalibard,Molmer,Carmichael,Dum}. This evolution consists of
two elements: evolution with a non hermitian hamiltonian, and
randomly decided ``quantum jumps", followed by wave function
normalization. This approach, which is equivalent to the master
equation treatment, has two main interests. First, if the relevant
Hilbert space of the quantum system has a dimension $N$ large
compared to 1, the number of variables involved in a wave function
treatment ($\sim N$) is much smaller than the one required for
calculations with density matrices ($\sim N^2$). Second, new
physical insight may be gained, in particular in the studies of the
behavior of a single quantum system.

\section {The MCWF procedure}

The class of relaxation operators that we consider here is the
following:
\begin{equation}
\lrel(\rho_S)=-{1\over 2}\sum_m \Big( C_m^\dagger C_m \rho_S+
\rho_SC_m^\dagger C_m \Big) + \sum_m C_m \rho_S C_m^\dagger \ .
\label{2.1}
\end{equation}
This type of relaxation operators is very general and it is found in
most of the Quantum Optics problems involving dissipation. In
(\ref{2.1}), the $C_m$'s are operators acting in the space of the
small system. Depending on the nature of the problem there can be
one, a few or an infinity of these operators.

For the particular case of spontaneous emission by a two-level
system with one stable ground state $g$ and one excited state $e$
with a lifetime $\Gamma^{-1}$, there is just a single operator
$C_1=\sqrt{\Gamma}|g\rangle \langle e|$ in the relaxation operator
(\ref{2.1}), and one can check that (\ref{2.1}) indeed leads to the
well known relaxation part of the optical Bloch equations:
\begin{equation}
\left\{
\begin{array}{ccr}
\dot{(\rho_S)}_{ee}& = &-\Gamma (\rho_S)_{ee} \\
\dot{(\rho_S)}_{gg}&= &\Gamma (\rho_S)_{ee}
\end{array}
\right.
\ \ \ \ \ \
\left\{
\begin{array}{ccr}
\dot{(\rho_S)}_{eg}& = &-(\Gamma/2) (\rho_S)_{eg} \\
\dot{(\rho_S)}_{ge}& = &-(\Gamma/2) (\rho_S)_{ge} \ .
\end{array}
\right.
\label{2.4}
\end{equation}

We now present the procedure for evolving wave functions of the
small system. Consider at time $t$ that the system is in a state
with the normalized wave function $|\phi(t)\ra$. In order to get the
wave function at time $t+\dt$, we proceed in two steps:

\begin{enumerate}

\item First we calculate the wave function $|\phia (t+\dt)\ra$
obtained by evolving $|\phi(t)\ra$ with the non hermitian Hamiltonian:
\begin{equation}
H=H_S-{i\hbar \over 2} \sum_m C_m^\dagger C_m \ .
\label {2.6}
\end{equation}
This gives for sufficiently small $\dt$:
\begin{equation}
|\phia (t+ \dt)\ra = \left(1-{iH\dt \over \hbar}\right)|\phi(t)\ra \ .
\label {2.7}
\end{equation}
Since $H$ is not hermitian, this new wave function is clearly not
normalized. The square of its norm is:
\begin{eqnarray}
\la \phia (t+ \dt)| \phia (t+ \dt)\ra & = & \la \phi(t)|
\left(1+{iH^\dagger\dt \over \hbar}\right)
\left(1-{iH\dt \over \hbar}\right) |\phi(t)\ra  \nonumber\\
&=& 1- \delta p
\label {2.8}
\end{eqnarray}
where $\delta p$ reads:
\begin{eqnarray}
\delta p & = &\dt\ {i \over \hbar}\ \la \phi(t) | H-H^\dagger
| \phi(t)\ra = \sum_m \delta p_m  \label {2.9a}\\
\delta p_m &=& \dt\ \la \phi(t) | C_m^\dagger C_m |\phi(t) \ra \
\ge \ 0 \ .
\label{2.9b}
\end{eqnarray}
The magnitude of the step $\dt$ is adjusted so that this calculation
at first order is valid; in particular it requires $\delta p \ll 1$.

For the particular case of the two-level atom problem, the non
hermitian Hamiltonian is
\begin{equation}
H=H_S- {i\hbar \Gamma \over 2} |e\ra \la e| \ .
\label{2.9c}
\end{equation}
This amounts to adding the imaginary term $-i\hbar \Gamma /2$ to the energy
of the unstable excited state, as usual in scattering theory.

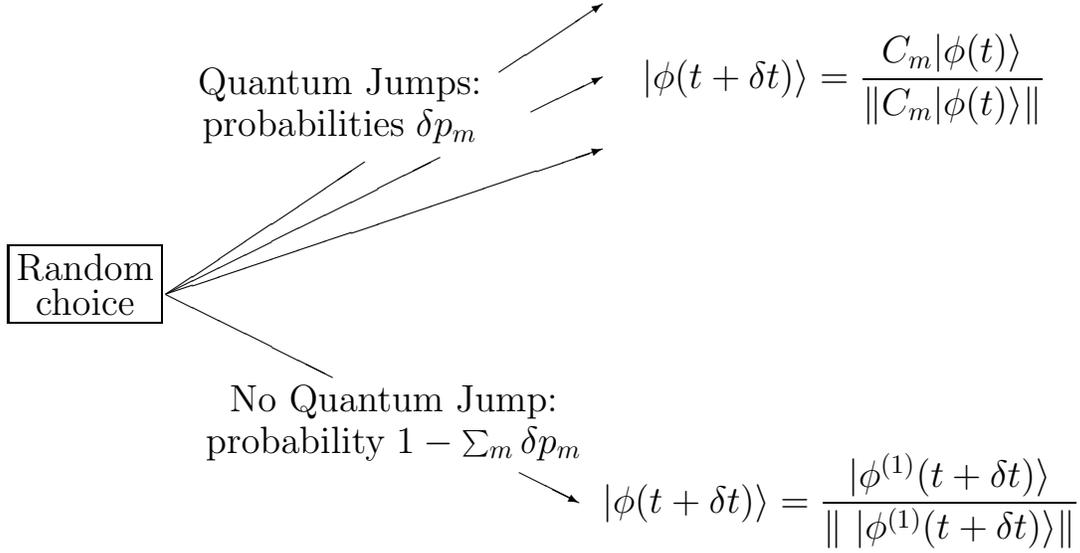
\begin{figure}[ht]
\unitlength=1.5pt
\begin{picture}(250,170)(5,30)
\large
\put (50,100){\line(3,2){50}}
\put (134,156){\vector(3,2){26}}
\put (50,100){\line(2,1){69}}
\put (142,146){\vector(2,1){18}}
\put (50,100){\vector(3,1){110}}
\put (50,100){\line(2,-1){42}}
\put (139,55){\vector(2,-1){15}}
\put (10,95){\fbox{\shortstack{Random \\ choice}}}
\put (58,140){\shortstack{Quantum Jumps: \\ probabilities
$\delta p_m$}}
\put (60,60){\shortstack{No Quantum Jump:
\\ probability $1-\sum_m\delta p_m$}}
\put (170,152){\shortstack
{$\displaystyle
|\phi(t+\delta t)\rangle={C_m |\phi(t)\rangle
\over \| C_m |\phi(t)\rangle \|}$}}
\put (160,45){\shortstack{$\displaystyle
|\phi(t + \delta t) \rangle = { | \phi^{(1)}(t+\delta t) \rangle \over
\| \ |\phi^{(1)} (t+\delta t)\rangle  \|  }$}}
\end{picture}

\caption{\sl The possible quantum jumps in the Monte-Carlo evolution}
\end{figure}

\item
The second step of the evolution of $|\phi\ra$ between $t$ and
$t+\dt$ consists in a possible ``quantum jump" (Fig. 1). The various
possible ``directions" for the jumps are given by the $C_m$
operators, and the probability for making a jump in the ``direction"
of a particular $C_m$ is given by $\delta p_m$ given in
(\ref{2.9b}). The new normalized wave function after such a jump is
given by:
\begin{equation}
\mbox{with a probability } \delta p_m \ \ \ \ \ \
|\phi (t+\dt)\ra={C_m|\phi(t)\ra \over  \| C_m |\phi(t)\rangle \|} \ .
\label{2.10}
\end{equation}
Using (\ref{2.9a}), we find that the total probability for making a jump
is $\delta p$. In the no-jump case, which occurs then with a probability $1-
\delta p$, we take as new normalized wave function at time $t+\delta t$:
\begin{equation}
\mbox{with a probability } 1-\delta p=1-\sum_m\delta p_m  \ \ \ \ \ \
|\phi (t+\dt)\ra={|\phia(t+\dt)\ra \over \| \ |\phi^{(1)} (t+\delta t)
\rangle  \| }\ .
\label{2.11}
\end{equation}

\end{enumerate}

Consider again as an example the particular case of the spontaneous emission of
a two-level atom. The wave function at time $t$ can be written as:
\begin{equation}
|\phi(t)\ra = \alpha(t) |e\ra + \beta (t) |g\ra \ .
\label{2.12}
\end{equation}
Since there is a single $C_m$ operator in this case, there is only one
possible type of quantum jump. The probability for this quantum jump
is:
\begin{equation}
\delta p= \Gamma |\alpha|^2 \delta t
\label{2.13}
\end{equation}
and the wave function after the jump is simply
$|\phi(t+\delta t)\ra=|g\ra$. If no jump occurs, the wave function at
time $t+\delta t$ is similar to (\ref{2.12}), with the coefficients
$\alpha(t+\delta t)$ and $\beta(t+\delta t)$ deduced from $\alpha(t)$
and $\beta(t)$ using the evolution with the non hermitian hamiltonian
(\ref{2.9c}). We see for this particular case that the
Monte-Carlo evolution can be understood as the
stochastic evolution of the atomic wave function if a continuous detection
of the emitted photons is performed. The probability for detecting a
photon during a particular time step $\delta t$ is indeed equal to
$\delta p$ given in (\ref{2.13}), and the new wave function after the
detection, according to the standard quantum measurement theory,
corresponds to the atom in its ground state $g$.

It is actually quite a general result that the Monte-Carlo evolution
outlined above represents
a possible history of the system wave function with a suitable
continuous detection
process taking place \cite{Dalibard,Carmichael}.
Although this procedure
does not make any reference to measurements on the system, it may be
useful, in order to get some physical understanding for the result of the
simulation, to refer to such a continuous detection process, as if it was
really performed.
We note in this respect that one might possibly consider several different
continuous detection
processes for a given quantum system. The various sets of $C_m$'s associated
to each of these detection schemes can be deduced from each other by
linear combinations,
the relaxation equation (\ref{2.1}) remaining then of course
unchanged~\cite{Molmer}.

\section{Equivalence with the Master Equation}

With this set of rules  we can propagate a wave function $|\phi(t)\ra$
in time, and we now show that this
procedure is equivalent to the master equation (\ref{relax}). More
precisely we consider the quantity $\bar \sigma (t)$ obtained by
averaging $\sigma(t)=|\phi(t)\ra \la \phi(t)|$ over the various
possible outcomes at time $t$ of the MCWF evolutions all starting in
$|\phi(0)\ra$, and we prove that $\bar \sigma(t)$ coincides with
$\rho_S(t)$ at all times $t$, provided they coincide at $t=0$.

Consider a given MCWF $|\phi(t)\ra$ at time $t$. At time $t+\dt$, the average
value of $\sigma (t+\dt)$ is:
\begin{eqnarray}
 \overline{\sigma(t+\dt)}=& &(1-\delta p)\ {|\phia(t+\dt)\ra \over
\| |\phi^{(1)} (t+\delta t)\rangle  \| }
{\la\phia(t+\dt)|\over \| |\phi^{(1)} (t+\delta t)\rangle  \| }
\nonumber \\
 &+& \sum_m \delta p_m {C_m |\phi(t)\ra \over
\| C_m |\phi(t)\rangle \| }{\la\phi(t)|C_m^\dagger\over
\| C_m |\phi(t)\rangle \| }
\label{2.14}
\end{eqnarray}
which gives, using (\ref{2.7}),(\ref{2.8}) and (\ref{2.10}):
\begin{equation}
\overline{\sigma(t+\dt)}= \sigma(t) + {i \dt \over \hbar} [\sigma(t),
H_S] +\dt \ \lrel(\sigma(t)).
\label{2.15}
\end{equation}
We now average this equation over the possible values of $\sigma(t)$
and we obtain:
\begin{equation}
{{\rm d} \bar \sigma \over {\rm d}t}={i\over \hbar}[\bar \sigma,H_S]
+ \lrel (\bar \sigma).
\label{2.15b}
\end{equation}
This equation is identical to the master equation (\ref{relax}). If
we assume that $\rho_S(0)=|\phi(0)\ra\la \phi(0)|$, $\bar \sigma(t)$
and $\rho_S(t)$ coincide at any time, which demonstrates the
equivalence between the two points of view. In the case where $\rho_S
(0)$ does not correspond to a pure state, one has first to decompose
it as a statistical mixture of pure states, $\rho(0)=\sum p_i |\chi_i
\ra\la \chi_i|$ and then randomly choose the initial MCWFs among the
$|\chi_i\ra$ with the probability law $p_i$.

As mentioned in the introduction, the master equation approach and the
reduced density matrix give access to  one time average values
 $a(t)=\la A\ra(t)={\rm Tr}\big(\rho_S(t)A\big)$, which can now
also be obtained with the MCWF method.
One calculates, for several outcomes
$|\phi^{(i)}(t)\ra$ of the MCWF evolution,
the quantum average $\la\phi^{(i)}(t)|A|
\phi^{(i)}(t)\ra$, and one takes the mean value of
this quantity over the various
outcomes $|\phi^{(i)}(t)\ra$:
\begin{equation}
\la A \ra_{(n)}(t)={1 \over n} \sum_{i=1}^n \la \phi^{(i)}(t)|A|
\phi^{(i)}(t)\ra \ .
\label{2.15c}
\end{equation}
For $n$ sufficiently large, (\ref{2.15b}) implies that
$\la A \ra _{(n)}(t) \simeq \la A \ra (t)$.
The ability to
provide expectation values for any system operator makes the
MCWF method a computational tool which may be much more efficient
than the numerical solution of (1) \cite{Molmer,Dum}.

As an example of the agreement between the master equation approach and
the MCWF approach, we have calculated by both methods the excited
state population of a two-level atom coupled to a coherent laser field.
The parameters for this Rabi nutation are a zero detuning $\delta$ between the
laser and atomic frequencies, and a Rabi frequency $\Omega=3\Gamma$.
In Fig. 2a, we show the excited state population for a single ``history"
for $|\phi(t)\ra$. One finds a continuous evolution for this
population oscillating between 0 and 1, interrupted by random
quantum jumps
projecting the atomic wave function into the ground state.
In Fig. 2b, we indicate the MCWF result obtained with the average of 100
wave functions. It shows a damped oscillation as a result of the
dephasing of the individual oscillations due to the randomness of the various
quantum jumps. The MCWF result is in good agreement with the one derived
from the master equation (Optical Bloch Equations). Note that the
purpose of this example is to illustrate the convergence of the two
methods, and not to provide an efficient way of treating two-level atom
problems. For such a small system, there is of course no gain in
computing time by using the MCWF method instead of the master equation.

\begin{figure}[htb]
 \epsfbox{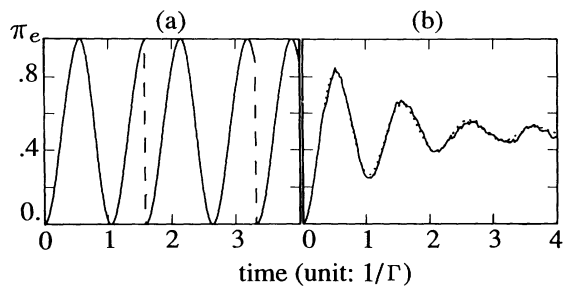}\\
  \caption{\sl (a) Time evolution of the excited-state population of a
two-level atom in the MCWF approach. The dashed lines indicate the projection of the atomic wave
function onto the ground state (quantum jump). (b) Excited state population averaged over 100 MCWF
starting all in ground state at time 0. The dotted line represents the master equation result.}
\end{figure}

It appears clearly that the equivalence of the Master
Equation and  MCWF approaches does not depend on the particular
value of the time step $\dt$. From a practical point of view,
the largest possible $\dt$ is preferable, and one might benefit
from using a generalization of (\ref{2.7}) to a higher
order in $\dt$, as for example a 4th order Runge-Kutta type
calculation.  The only requirement on $\dt$ is that the various
$|\eta_i|\dt$, where the $\hbar \eta_i$ are the eigenvalues of $H$,
should
be small compared to 1. Of course we assume here that those
eigenvalues have been simplified as much as possible in order to
eliminate the bare energies of the eigenstates of $H_S$. For instance,
for a  two-level atom with a transition frequency $\omega_A$
coupled to a laser field with frequency $\omega_L$, one makes
the rotating wave approximation in the rotating frame
 so that the $|\eta_i|$'s are of the order
of the natural width $\Gamma$, the Rabi frequency $\Omega$ or
the detuning $\delta=\omega_L-\omega_A$; they are
consequently much smaller than $\omega_A$.

One might wonder whether there is a minimal size for
the time step $\dt$. In
the derivation presented above, it can be  chosen arbitrarily
small. However one should remember that the
derivation of (\ref{relax}) involves a coarse grain average of
the real density operator evolution. The time step of this
coarse grain average has to be much larger than the correlation time
$\tau_c$ of the reservoir, which is typically an optical period
for the problem of spontaneous emission. Therefore one should be
cautious when considering any result derived
from this MCWF approach involving details with a time scale of
the order of or shorter than $\tau_c$, and only $\dt$ larger than
$\tau_c$ should be applied.
This appears clearly if one starts directly from the interaction
Hamiltonian between the system and the reservoir in order to generate
the stochastic evolution for the system wave function
\cite{Dalibard}. The condition $\dt \gg \tau_c$ is then required to
prevent quantum Zeno type effects \cite{Misra}.
This restriction is discussed
in detail in \cite{Hegerfeldt} in connection with quantum measurement
theory.

\section{MCWF and other stochastic approaches}

The problem of stochastic wave function evolution in connection with the
treatment of dissipative systems in quantum optics
has recently received a lot of attention.
In the context of non classical field generation,
Carmichael \cite{Carmichael} has proposed an approach named
``quantum trajectories", inspired by the theory of photoelectron
counting sequences \cite{Kelley} and quite similar to the spirit of
the present work.

For simple atomic systems (2 or 3 levels) coupled to the electromagnetic
field, the dynamics can be interpreted
in terms of one or a few {\em delay functions}, which give the probability
distribution of the time intervals between the emission of two successive
photons \cite{Cohen2,Zoller,Carmichael2}.
When these functions are known analytically, they
can generate a very efficient Monte-Carlo analysis of the process: just after
the emission of the $n^{th}$ fluorescence photon at time $t_n$,
the atom is in its ground
state and the choice of a single random number is sufficient
to determine
the time $t_{n+1}$ of emission of the $n+1^{st}$ photon. This type of
Monte-Carlo analysis has been used in \cite{Blatt} to
simulate an atomic beam
cooling experiment, and in \cite{Zoller} to prove
numerically the existence
of dark periods in the fluorescence of a 3-level atom (quantum jumps).
Very recently, laser cooling of atoms using velocity selective coherent
population trapping \cite{Cohen3} and lasing without inversion
\cite{Cohen4} have been analyzed by this type of
Monte-Carlo method.

Unfortunately, the delay function cannot be calculated analytically
for complex systems involving a large number of levels.
Nevertheless, it is possible to generate a Monte-Carlo solution for this
problem in which a single random number determines the time of
emission of each fluorescence photon \cite{Dum}.
The evolution of the system between two quantum jumps has to be integrated
step by step
numerically, so that the amount
of calculation involved is similar to the one
required by the method presented in this paper.

Stochastic approaches have also been introduced in the context of either
standard \cite{Meystre87,Ueda90} or quantum
non demolition \cite{Brune90,Haroche92,Ueda92}
measurements of photon number in a
given mode of the electromagnetic field.
A sequence of random quantum jumps resulting from successive measurements
asymptotically
leads to a reduction of the field state into a Fock state $|n\ra$, whose
probability distribution is equal to the initial photon number distribution
for the case of the non demolition measurement.
The main interest of these stochastic approaches, as compared with the usual
master equation treatment, is to give explicit individual histories of the
quantum field state in a measurement sequence. This is particularly
valuable if one wants to optimize the measurement sequence in order
to get a complete information on the field state with a minimum number
of measurement processes \cite{Haroche92}.
On the other hand, these stochastic calculations
still mostly deal with density matrices and their authors do not seem to
consider them as more efficient ways of computing than the master equation.

Another class of stochastic equations for system wave functions,
which is also equivalent to the master equation (\ref{relax}), has been
introduced by Gisin and Percival \cite{Gisin}
(see also the work by Diosi \cite{Diosi}). In this approach, only
continuous stochastic equations are considered. The complex
It\^o stochastic process is given by:
\begin{eqnarray}
|d\phi\ra = -{i\over \hbar} H_S dt
|\phi\ra &+& \sum_m \Big( \la C^\dagger_m \ra
C_m-{1\over 2}C_m^\dagger C_m - {1\over 2} \la C_m^\dagger \ra \la C_m \ra
\Big) |\phi\ra dt \nonumber \\
&+& \sum_m\Big( C_m - \la C_m\ra \Big) |\phi\ra \ {d\xi_m \over \sqrt 2}
\label{3.1}
\end{eqnarray}
where $\la C_m\ra=\la \phi |C_m |\phi \ra$,
and where the $d\xi_m$ are independent
complex Wiener processes \cite{Gardiner}:
\begin{eqnarray}
&&\overline{d\xi_m}=0 \nonumber \\
&&\overline{\Re(d\xi_m) \Re (d\xi_n)} = \overline{\Im (d\xi_m) \Im (d\xi_n)}
=\delta_{m,n} dt  \label{3.2} \\
&&\overline{\Re (d\xi_m) \Im (d\xi_n)} =0 \nonumber
\end{eqnarray}

\def\e{\varepsilon}
\def\un{\rlap{\small 1}\kern .15em 1}
Carmichael has shown that for the particular case of the homodyne detection of
the fluorescence light,
the Quantum Jump formalism can be transformed into such a
continuous stochastic equation \cite{Carmichael}.
Actually this proof can be extended to the
most general case:
the first step is to write the relaxation operator $\lrel$ as:
\begin{equation}
\lrel(\rho_S)=-{1\over 2} \sum_{m,\e} \Big( D^\dagger_{m,\e}D_{m,\e}\ \rho_S
+ \rho_S\ D^\dagger_{m,\e}D_{m,\e} \Big) + \sum_{m,\e} D_{m,\e}\ \rho_S\
D_{m,\e}^\dagger
\label{3.3}
\end{equation}
where $\e=\pm 1$ and where the $D_{m,\e}$ are defined as:
\begin{equation}
D_{m,\e}={\mu \un + \e C_m \over \sqrt{2}}
\label{3.4}
\end{equation}
One easily shows that $\lrel$ in (\ref{3.3}) is identical with the one in (2). The coefficient $\mu$
is arbitrary at this stage; $\mu^2$ has the dimension of the inverse of a time, and we just require in
the following $\mu^2 \gg |\eta|$, where $\hbar \eta$ is a typical eigenvalue for $H$ (for the
two-level atom case, $\eta \sim \Gamma,\Omega,\delta$). Using the set of operators $D_{m,\e}$, we can
now perform a Monte-Carlo evolution of the wave function, equivalent to the master equation
(\ref{relax}). Because of the large magnitude of $\mu^2$, this simulation with the $D_{m,\e}$
operators involves much more quantum jumps in a given time interval $\Delta t$ than a simulation done
with the $C_m$'s. But the change of the wave function in a given quantum jump:
\begin{equation}
| \phi \ra \longrightarrow {D_{m,\e} |\phi \ra \over \| D_{m,\e}|\phi\ra \| }
\label{3.5}
\end{equation}
is very small since $D_{m,\e}$ is nearly proportional to the identity operator
$\un$.
In the limit of very large $\mu$, the Monte-Carlo evolution of the
wave function therefore tends towards a continuous stochastic evolution.
In Carmichael's homodyne detection problem, the form
(\ref{3.4}) for the $D_{m,\e}$ has a clear interpretation. These jump operators
correspond to the detection of a photon after one has mixed the light emitted
by the atomic system with a local oscillator field. The parts in $\mu \un$ and
$C_m$ correspond respectively to the field originating from the local
oscillator and the field emitted by the atom. The condition $\mu^2 \gg |\eta|$
just states that the intensity of the local oscillator is much larger than the
intensity of the light emitted by the atom, as usual in homodyne detection.

To prove the equivalence with (\ref{3.1}),
we choose a time interval $\Delta t$ such that
\begin{equation}
\mu^{-2} \ll \Delta t \ll |\eta|^{-1}\ .
\label{3.6}
\end{equation}
This implies that the number of jumps $N_{m,\e}$ occurring
with a given operator $D_{m,\e}$
during $\Delta t$ will be large compared to 1  since $ \mu^2\ \Delta t \gg 1$,
but at the same time we expect
only a small change in the system wave function since $|\eta| \Delta t \ll 1$.
The operator ${\cal O}$ describing the action of all
those jumps during $\Delta t$
is a product of the various $D_{m,\e}$ and it can be approximated at order 1
in $\sqrt{\Delta t\ |\eta|} $ by:
\begin{equation}
{\cal O} \simeq
\left({\mu \over \sqrt{2}}\right)^N \Big( \un + {1\over \mu} \sum_m
(N_{m,+}-N_{m,-})\ C_m \Big)
\label{3.7}
\end{equation}
where $N=\sum_{m,\e} N_{m,\e}$ is the total number of jumps occurring during
$\Delta t$.
The wave function at time $t+\Delta t$ can now be written before
normalization:
\begin{equation}
|\phi(t+\Delta t)\ra = \Big( \un  -{i\over \hbar}H_S \Delta t
-{1 \over 2} \Delta t \sum_m C_m^\dagger C_m
+ \sum_m {N_{m,+}-N_{m,-} \over \mu } C_m \Big) |\phi(t)\ra
\label{3.8}
\end{equation}
where we have taken into account both the non-hermitian evolution during
$\Delta t$ and the effect of the multiple quantum jumps.
The numbers of jumps $N_{m,\e}$ are poissonian random variables with an
average value and a standard deviation given by:
\begin{eqnarray}
\overline {N_{m,\e}}
& \simeq & {\mu^2 \Delta t \over 2} \left( 1 + {\e \over \mu}
\la C_m+ C_m^\dagger \ra \right)
\label{3.9} \\
\Delta N_{m,\e} & \simeq & {\mu \over \sqrt 2} \sqrt{\Delta t}
\label{3.10}
\end{eqnarray}
where the average value $\la C_m+C_m^\dagger \ra$ is taken in $|\phi(t)\ra$.
In the limit of large $N_{m,\e}$, we can approximate
the random variable $N_{m,+}-N_{m,-}$ appearing in
(\ref{3.8}) by:
\begin{equation}
{N_{m,+}-N_{m,-} \over \mu}= \Delta t \la C_m + C_m^\dagger \ra
+ \Delta \zeta_m
\label{3.11}
\end{equation}
where $\Delta \zeta_m$ is a real gaussian random variable with zero mean and a
standard deviation equal to $\sqrt{\Delta t}$.
Finally we normalize the wave function (\ref{3.8}) and we obtain:
\begin{eqnarray}
|\Delta \phi(t+\Delta t)\ra =  &- &{i\over \hbar} H_S |\phi(t)\ra \ \Delta t
\nonumber \\
&+& {1 \over 2}
\sum_m \left( \la C_m + C_m^\dagger \ra
C_m- C_m^\dagger C_m - {1 \over 4}
\la C_m + C_m^\dagger\ra^2
\right)|\phi(t)\ra \Delta t \nonumber \\
&+& {1 \over 2} \sum_m \left(2 C_m - \la C_m + C_m^\dagger \ra \right)
|\phi(t)\ra
\Delta \zeta_m \ .
\label{3.12}
\end{eqnarray}
In (\ref{3.12}), we have kept terms linear in $\Delta \zeta_m$ and $\Delta t$,
and we have replaced all the quadratic terms $\Delta \zeta_m \Delta \zeta_{m'}$
by their mean $\Delta t\ \delta_{m,m'}$.
In the limit $\mu \rightarrow +\infty$,
$\Delta t \rightarrow 0$, this equation can be understood as a It\^o
stochastic equation, corresponding to a real version of (\ref{3.1}).

The exact
form of (\ref{3.1}) can be recovered by taking a slightly more complicated
set of $D_{m,\e}$ operators:
\begin{equation}
D_{m,\e}={\mu \un + \e C_m \over 2} \ \ \ \ \ \ \mbox{with} \  \ \
\e=\pm 1,\pm i
\label{3.13}
\end{equation}
and by performing an appropriate global phase change of the wave function.
The continuous stochastic equation (\ref{3.1}) is therefore a limiting
case of the quantum jump formalism presented here, and it also has an
interpretation in terms of
a detection scheme: the information concerning the system is mixed
with a ``classical field" $\mu \un$,
and the sequence of quantum jumps deduced from
the whole set of mixed components
$D_{m,\e}$ allows one to determine the subsequent system evolution.
Note that on the contrary, the jumps deduced from the action
on the system wave function of a single
mixed component, such as the $D_{m,+}$'s,
are not sufficient to determine this system evolution.

\section{Conclusion}
We have presented a stochastic evolution for the wave function of a system
coupled to a reservoir in the Markovian regime.
Each time step in this stochastic evolution consists in two parts: a
Hamiltonian but non
hermitian evolution and a possible quantum jump.
We have proved the equivalence
of this Monte-Carlo Wave function approach with the master equation treatment.
We have also shown that this simulation with Quantum jumps can be
transformed into a continuous stochastic evolution of the wave function,
similar to the one of \cite{Gisin}.

This approach provides a computational tool which is often more efficient
than the standard master equation treatment for systems with a number of
states $N\gg 1$ (for a detailed discussion see \cite{Molmer}). Indeed a wave
function involves only $N$ components while a density matrix is described by
$N^2$ terms. This method has already been applied successfully
to problems such as
the study of the limits of laser cooling in 2 dimensions \cite{Berg},
or the calculation
of the spectrum of the light emitted by an assembly of cold atoms
\cite{Marte}. Problems such as the study of collisions between cold atoms,
or non linear mixing of quantum fields may also benefit from such an approach.

We have emphasized that this simulation is in many practical cases directly
connected to a measurement sequence performed on the system. Each
Monte-Carlo trajectory is a possible history for the individual quantum system.
In this respect, the noise appearing
when one simulates with this method
the measurement of a given observable $A$ is also
interesting. The fluctuations in the number of occurrences of a given eigenvalue
$a_i$ of $A$ correspond to the quantum noise that one would get
in a real experiment, performing the relevant detection scheme on an
individual quantum system. Since more and more quantum optics and atomic
physics experiments are now performed with a single system (single ion or
atom, single mode of a cavity), Monte-Carlo wave function
methods should therefore have many applications, since they
lead to predictions closer to
actual experimental signals than the master equation,
which rather deals with ensemble averages.

\end{document}